\documentclass{article}
\usepackage{fortschritte}
\usepackage[centertags]{amsmath}
\usepackage{amsfonts} \usepackage{amssymb} \usepackage{amsthm}
\def\beq{\begin{equation}}                     % 
\def\eeq{\end{equation}}                       %
\def\bea{\begin{eqnarray}}                     %         %
\def\eea{\end{eqnarray}}                       %       % 
\begin {document}                 
\rightline{NORDITA-2004-112} \rightline{DSF-45/2004}

\def\titleline{
%%%%%%%%%%%%%%%%%%%%%%%%%%%%%%%%%%%%%%%%%%%%%%
%                                            
% Insert now the text of your title.         
% Make a linebreak in the title with         
%                                            
%            \newtitleline                   
%                                            
%%%%%%%%%%%%%%%%%%%%%%%%%%%%%%%%%%%%%%%%%%%%%%%
The Gauge/Gravity Correspondence\\                        %         %
for Non-Supersymmetric Theories                                 %       %
%                                             %     %%%%%%%%%%%%%%
%                                             %       %
%%%%%%%%%%%%%%%%%%%%%%%%%%%%%%%%%%%%%%%%%%%%%%%         %
}
\def\email_speaker{
{\tt 
%%%%%%%%%%%%%%%%%%%%%%%%%%%%%%%%%%%%%%%%%%%%%%
%                                                  
% Insert now the e-mail address of the speaker or  
% the author that should get the electronic mail   
% of the publishing house                           
%                                                  
%%%%%%%%%%%%%%%%%%%%%%%%%%%%%%%%%%%%%%%%%%%%%%         %
divecchi@alf.nbi.dk             %       %
%                                            %     %%%%%%%%%%%%%%
%                                            %       %       
%%%%%%%%%%%%%%%%%%%%%%%%%%%%%%%%%%%%%%%%%%%%%%         %
}}
\def\authors{
%                                            
%  Insert now the name (names) of the author 
%  (authors).                                
%  In the case of several authors with       
%  different addresses use labels e.g.       
%                                            
%             \1ad  , \2ad  etc.             
%                                            
%  to indicate that a given author has the   
%  address number 1 , 2 , etc.               
%  Signify the person with the above e-mail  
%  address with \sp (behind the address      
%  labels, e.g.                              
%                                            
%            \2ad\sp                         
%                                            
%  when the 2nd author                       
%  happens to be the 'speaker')                                  
%%%%%%%%%%%%%%%%%%%%%%%%%%%%%%%%%%%%%%%%%%%%%%         %
P. Di Vecchia\1ad\sp , A. Liccardo\2ad ,           %       %
R. Marotta\2ad and F. Pezzella\2ad                        %     %%%%%%%%%%%%%
%                                            %       %
%%%%%%%%%%%%%%%%%%%%%%%%%%%%%%%%%%%%%%%%%%%%%%         %
}
\def\addresses{
%%%%%%%%%%%%%%%%%%%%%%%%%%%%%%%%%%%%%%%%%%%%%%
%                                            
% List now the address. In the case of       
% several addresses list them by numbers     
% using e.g.                                 
%                                            
%             \1ad , \2ad   etc.             
%                                            
% to numerate address 1 , 2 , etc.           
%                                            
%%%%%%%%%%%%%%%%%%%%%%%%%%%%%%%%%%%%%%%%%%%%%
\1ad                                        %          
Nordita, Blegdamsvej 17, 2100 Copenhagen {\O}, Denmark\\            %         %
\2ad                                        %       %
Dipartimento di Scienze Fisiche, Universit{\'{a}} di Napoli and INFN,
Sezione di Napoli \\%%
Complesso Universitario Monte S. Angelo, ed. G, via Cintia, 80126
Napoli, Italy\\           %       %
%\3ad                                        %         %
%Institute Three, University Three \\        %          
%Road Three, Town Three, Contry Three\\      %
%%%%%%%%%%%%%%%%%%%%%%%%%%%%%%%%%%%%%%%%%%%%%
}
\def\abstracttext{
%%%%%%%%%%%%%%%%%%%%%%%%%%%%%%%%%%%%%%%%%%%%%
%                                             
% Insert now the text of your abstract.       
%                                             
%%%%%%%%%%%%%%%%%%%%%%%%%%%%%%%%%%%%%%%%%%%%%         %
We review the string construction of the ``orientifold field
theories'' and we show that for these theories the gauge/gravity 
correspondence is only valid for a large number of colours. \\                 
%       %                                        %     %%%%%%%%%%%%%%%%%
%                                           %       %  
%%%%%%%%%%%%%%%%%%%%%%%%%%%%%%%%%%%%%%%%%%%%%         %
}
\large
\makefront
%%%%%%%%%%%%%%%%%%%%%%%%%%%%%%%%%%%%%%%%%%%%%%%%
%                                              %
%  Insert now the remaining parts of           %
%  your article.                               %
%                                              %
%%%%%%%%%%%%%%%%%%%%%%%%%%%%%%%%%%%%%%%%%%%%%%%%
\section{Introduction}

In recent years it has been possible to get a lot of perturbative and
non-perturbative information about less supersymmetric and
non-conformal four-dimensional gauge theories living on the
world-volume of wrapped and fractional D branes by using their 
corresponding classical solutions of the equations of motion of 
the low-energy string effective action~\footnote{For general reviews
  of various approaches see Ref.s~\cite{REV,KLEBA,MA,MILA,DL0307}.}. 
These equations contain not only
the supergravity fields present in the bulk ten-dimensional action but
also boundary terms corresponding to the location of the branes. It
turns out that in general  the classical solution develops  a
naked singularity of the repulson type at short distances from the
branes. On the other hand at these distances it does not provide
anymore a reliable
description of the branes because of the presence of an enhan{\c{c}}on
located at distances slightly higher than the naked 
singularity~\cite{enhancon}. The
enhan{\c{c}}on radius corresponds in the gauge theory living on the branes
to the dynamically generated scale $\Lambda_{QCD}$. Then, since short 
distances in supergravity correspond to large distances in the gauge
theory, as implied by holography, the presence of the enhan{\c{c}}on
does not allow  to get information about the large distance
behaviour of the gauge theory living on the D branes.  Above the
radius of the enhan{\c{c}}on instead the classical solution provides a good
description of the branes and therefore it can be used to get information
on the perturbative behaviour of the gauge theory. This is why 
the perturbative behaviour of ${\cal{N}}=2$ super Yang-Mills with
gauge group $SU(N)$ has been obtained from the classical solution 
corresponding to a bunch of $N$ fractional D3 branes of the orbifold
${\mathbb{C}}^2 / {\mathbb{Z}}_{2}$~\cite{KLENE,d3,polch}. On 
the other hand, from the point of
view of string theory, the one-loop perturbative behaviour of the gauge
theory living on the world-volume of a system of D branes can be
obtained from the annulus diagram by performing the
field theory limit in the open string channel. This is a consequence
of the fact that this limit has the
effect of allowing  only the massless open string states to
circulate in the loop. But, since the classical solution involves only
the massless closed string states, the question is then: why is the 
contribution of
the massless open string states  equivalent to that of the massless
closed string states? The previous results seem to imply that the
contribution of the massless open string states in the annulus
diagram is transformed under open/closed string duality into that of
the massless closed string states. This has been shown~\cite{LMP} in fact 
to happen for the
fractional D3 branes of the orbifolds $\mathbb{C}^2 / \mathbb{Z}_{2}$
and $\mathbb{C}^3 / (\mathbb{Z}_{2} \times \mathbb{Z}_{2}) $ of type
IIB that
describe respectively ${\cal{N}}=2$ and ${\cal{N}}=1$ super
Yang-Mills. In particular, in these cases it has been shown that the
threshold corrections due to the massive string states are identically
vanishing. These properties seem to be valid also for 
fractional D3 branes on the conifold and for D branes wrapped on
Calabi-Yau two-cycles where an explicit string analysis is not
possible.  Can these properties be extended to non-supersymmetric
theories? Recently the so called ``orientifold field theories'' have
been studied. They have the property of reducing to a supersymmetric
one in the large number $N$ of colours. A string description of some  
of them has been provided in Ref.~\cite{DLMP0407} where  it has been also
analyzed if the properties of the supersymmetric theories discussed
above  can be extended to them. In this talk we will review these
results. In Sect. (\ref{sect2}) we summarize what happens in the case
of the supersymmetric ${\cal{N}}=2$ theories. In Sect. (\ref{sect3})
we give a string construction of some orientifold field theories and
in Sect. (\ref{sect4}) we analyze if the gauge/gravity correspondence
is valid for them.

\section{Gauge/gravity correspondence for supersymmetric theories}
\label{sect2}

Let us consider $N$ fractional D3 branes of type IIB string theory on 
the orbifold   ${\rm I\!R}^{1,5}\times \mathbb{C}^2/\mathbb{Z}_2$
having their world-volume along the directions $0, 1, 2, 3$, while the
non-trivial element of $\mathbb{Z}_2 $  acts as $ x^i \rightarrow -x^i$
along the directions $i=6, 7, 8, 9$. 
The fractional D3 branes  are coupled to the untwisted massless
closed string excitations  corresponding to the graviton, the dilaton
and the  R-R $4$-form $C_4$ and to the twisted ones corresponding
to $ B_2 = \omega_2 b$ and $ C_2 = \omega_2 c$,   
where $\omega_2$ is the volume-form of the vanishing $2$-cycle located
at the orbifold fixed point and such that $\int_{{\cal{C}}_2} \omega_2
=1$. The classical supergravity solution
corresponding  to a fractional D3 brane is given by 
($\alpha,\beta=0\dots 3$, $i,j=4\dots 9$):
\begin{eqnarray}
&&ds^2 = H^{-1/2}\, \eta_{\alpha\beta}\,d x^\alpha dx^\beta +
H^{1/2} \,\delta_{ij} \,dx^i dx^j~~,
\label{met48} \\
&&{\widetilde{F}}_{(5)} =  d \left(H^{-1} \, dx^0 \wedge \dots
\wedge dx^3 \right)+ {}^* d \left(H^{-1} \, dx^0 \wedge \dots
\wedge dx^3 \right)~~, \label{f5ans}\\
&&c=-N(2 \pi \sqrt{\alpha'})^2  \frac{g_s}{\pi} \theta~~;~~
b= N (2 \pi \sqrt{\alpha'})^2  \frac{g_s}{\pi}\log(\rho/\epsilon) 
\label{bfin}
\end{eqnarray}
where $ z \equiv x^4 + i x^5 = \rho {e}^{i \theta}$ and $H$ is given in
Ref.~\cite{d3}. 

% && H = 1+ \frac{Q}{r^4}  + \frac{2 K^2}{r^4} \left[\log
%\Bigg(\frac{r^4}{\epsilon^2 (r^2-\rho^2)}\Bigg) - 1+ \frac{\rho^2
%}{r^2-\rho^2} \right]\nonumber \\
%&&\hspace{9.5cm}\Red{Q=\alpha' \,K}\nonumber
%\end{eqnarray}}
The supergravity solution provides a non-trivial
information on the four-dimensional gauge theory living on the
world-volume of $N$ fractional branes, namely  $SU(N)$ ${\cal{N}}=2$ super
Yang-Mills. This can be obtained by
considering  the world-volume theory (Dirac-Born-Infeld +
Wess-Zumino term) of a fractional D3 brane. 
In the $\alpha'\rightarrow 0$ limit,
keeping fixed the scalar field
$\phi=\frac{x_4+i\,x_5}{2\pi \sqrt{2}\alpha'}$ of the gauge
multiplet of  ${\cal N}=2$ super Yang-Mills, one obtains
\begin{eqnarray}
S_{\rm YM}= -\,\frac{1}{g^{2}_{\rm YM}} \int d^4 x  \left\{
\frac{1}{4}
 F_{\alpha \beta}^a F^{\alpha \beta}_a + \frac{1}{2} \partial_{\alpha}
 {\overline\phi}\, \partial^{\alpha} \phi \right\} + \frac{\theta_{\rm
 YM}}{32 \pi^2} \int d^4 x F_{\alpha \beta}^a {\widetilde{F}}^{\alpha
 \beta}_a 
\label{sym}
\end{eqnarray}
where the Yang-Mills coupling constant is given by:
\begin{eqnarray}
\frac{1}{g^{2}_{\rm YM}  } = \tau_5 \frac{(2 \pi \alpha')^2}{2}
\int_{{\cal{C}}_2}
%e^{- \varphi} 
B_2 = \frac{
%{\rm e}^{-\varphi}\,
b}{16\pi^3\alpha'g_s} = \frac{1}{8\pi g_s} +
\frac{2N}{8 \pi^2} \log \frac{\rho}{\epsilon}
\label{gym}
\end{eqnarray}
and the {$\theta$-angle} by ($ \tau_5 = [g_s \sqrt{\alpha'} (2 \pi
\sqrt{\alpha'})^5 ]^{-1}$):
\begin{eqnarray}
 \theta_{\rm YM} = \tau_5 (2 \pi \alpha')^2 (2 \pi)^2 \int_{{\cal{C}}_2} C_2
=  \frac{c}{2\pi\alpha'g_s} = - 2N
\; \theta . 
\label{thetaym}
\end{eqnarray}
Since the supergravity coordinate $z$ corresponds to the complex
scalar field $\phi$ of ${\cal{N}}=2$ super Yang-Mills and we know how
the scale and chiral anomalous transformations act on it, we can infer
how they act on $z$:
\begin{eqnarray}
\phi \rightarrow  \mu\,{\rm e}^{{2}\,{\rm i} \,\alpha} \phi
\Longleftrightarrow 
z \rightarrow \mu\,{\rm e}^{{2}\,{\rm i} \,\alpha} \,z .
\label{phiz}
\end{eqnarray}
Acting with the previous transformations in Eq.s (\ref{gym})
and (\ref{thetaym}) we get:
\begin{eqnarray}
\frac{1}{g^{2}_{\rm YM}} \rightarrow \frac{1}{g^{2}_{\rm YM}}
+\frac{2N}{8\pi^2}\,\log\mu~~~~~;~~~~~ \theta_{\rm YM}
\rightarrow \theta_{\rm YM} -4N\,\alpha~~. 
\label{gtheta}
\end{eqnarray}
These equations reproduce the $\beta$-function and the chiral anomaly
of $SU(N)$ ${\cal{N}}=2$ super Yang-Mills.

We will now show that the fact that the supergravity solution 
reproduces the perturbative
properties of the gauge theory living in the world-volume of $N$
fractional D3 branes is a consequence of the absence of threshold
corrections to the running gauge coupling constant and the
$\theta$-angle. This can be seen  by computing the annulus
diagram describing the interaction between a set of $N$ fractional
D3 branes and a fractional D3 brane having an external gauge field
in its world-volume and located at a distance $\rho$ from the
 other branes. In the case of the orbifold  
$\mathbb{R}^{1,5}\times\mathbb{C}^2/\mathbb{Z}_2$ the annulus diagram
has two contributions: one from the untwisted and the other from the
twisted sector:
\begin{eqnarray}
Z =\int_{0}^{\infty} \frac{d \tau}{\tau} Tr_{{\rm
    NS-R}} \left[ \left(\frac{1 +h}{2}\right)  (-1)^{G_{bc}} P_{GSO}
%\left(     \frac{(-1)^{G_{\beta \gamma}} + (-1)^F }{2} \right) 
\,\, { e}^{- 2 \pi \tau L_0} \right] =
Z_e  +Z_h~~,
\label{anulu}
\end{eqnarray}
but only the twisted sector gives a non-zero contribution to the
anomalies. One can extract from the annulus diagram
the contribution to the gauge coupling constant and to the
$\theta$ vacuum. It can be extracted either from the open or from
 the closed string channel.
From the open string channel one gets:
\begin{eqnarray} 
%Z_h^o(F)\!\!&\rightarrow&\!\!
\left[-
\frac{1}{4} \int d^4 x F^2 
\right] \left[ - \frac{N}{8 \pi^2}
\int_{1/(\alpha' \Lambda^2)}^{\infty} \frac{d \tau}{\tau}
{e}^{-\frac{\rho^{2} \tau}{2 \pi \alpha' } } \right]  
 - iN \left[\frac{1}{32\pi^2} \int d^4x F \cdot {\tilde{F}}
 \right] \int_{0}^{\infty}
\frac{d\tau}{\tau} e^{ -\frac{\rho^{2} \tau}{2\pi\alpha'} } .
\label{open21}
\end{eqnarray}
Notice that the sign in front of the topological term is opposite to the
sign appearing in Ref.s~\cite{LMP,DLMP0407} because in these papers we
have used a different GSO projection in the R sector.
From Eq. (\ref{open21}) we see that only the massless open string 
states contribute
to the gauge coupling constants and their contribution under
open/closed string duality goes exactly into that of the massless
closed string states that appear in the supergravity solution. The
massive states in both channels give no contribution and this implies
the absence of threshold corrections. This is basically the reason why
one can read the perturbative behaviour of ${\cal{N}}=2$ super
Yang-Mills from the classical supergravity solution. The expressions
that we get from Eq. (\ref{open21}) are, however, divergent already at
the string level and in addition the $\theta$-angle is imaginary. In
Ref.~\cite{DLMP0407} we have discussed how to cure the two problems by
introducing a complex cut-off. Here we give just the final results:
\begin{eqnarray}
\frac{1}{g_{YM}^{2}}= \frac{1}{g_{YM}^{2} (\Lambda)} - \frac{N}{8
  \pi^2}  \frac{1}{2} \left[ I (z) + I ({\bar{z}}) \right] = 
\frac{1}{g_{YM}^{2} (\Lambda)} + \frac{N}{8
  \pi^2} \log \frac{\rho^2}{2 \pi (\alpha')^2 \Lambda^2}
\label{coupli89}
\end{eqnarray}
and 
\begin{eqnarray}
\theta_{YM} = -i N \cdot \frac{1}{2} \left[ I (z) - I ({\bar{z}})
\right] = - 2 N \theta
\label{the89}
\end{eqnarray}
where
\begin{eqnarray}
I (z) \equiv \int_{1/(e^{-i \theta}\Lambda)^2}^{\infty}  \frac{d\sigma}{\sigma}
e^{-\frac{\rho^2 \sigma}{2\pi (\alpha')^2}} \simeq \log \frac{2 \pi
  (\alpha')^2 \Lambda^2}{\rho^2 {e}^{ 2i \theta}}~~~;~~ z\equiv \rho 
e^{i \theta}
\label{comple68}
\end{eqnarray}
and we have introduced a term corresponding to the bare coupling
constant $g_{YM} (\Lambda )$. Eq.s (\ref{coupli89}) and (\ref{the89})
are identical to Eq.s (\ref{gym}) and (\ref{thetaym}).

\section{The string construction of the ``orientifold field theories''}
\label{sect3}
 
In this section we summarize the string construction of the so called
``orientifold field theories''~\footnote{See Ref.~\cite{DLMP0407} for
details, References therein and Ref.~\cite{SAGNOTTI} for the original
construction of tachyon free  orientifolds of $0B$ theory. See also  
Ref.~\cite{ASV} for a review of the properties of these
  theories.} . We consider the orientifold  
$ \frac{0B}{\Omega' I_6 (-1)^{F_L}}$ of type 0B string theory and we 
consider a D3 brane of this theory. Unlike 0B, 
it is a fully consistent theory with no tachyons and no 
R-R and NS-NS tadpoles. It can be shown that
the gauge theory living on $N$ D3 branes of this orientifold is a
non-supersymmetric gauge theory with a gluon, six adjoint scalars and
four Dirac fermions in the two-index (anti)symmetric representation of
$SU(N)$. This gauge theory contains $8N^2$ bosonic and $8N(N\pm1)$
fermionic degrees of freedom. Its one-loop $\beta$-function is equal
to:
\begin{eqnarray}
\beta (g_{YM} ) = \frac{g_{YM}^{3}}{(4 \pi)^2} \left[- \frac{11}{3}N +
6 \cdot \frac{N}{6} + 4 \cdot \frac{4}{3} \frac{N \pm 2}{2} \right] =
\frac{g_{YM}^{3}}{(4 \pi)^2} \left[ 0 \cdot N \pm \frac{16}{3} \right].
\label{beta1}
\end{eqnarray}   
For large $N$ one recovers the Bose-Fermi degeneracy and the
$\beta$-function vanishes as the one of ${\cal{N}}=4$ super
Yang-Mills. We call it ``orientifold field theory'' because it is
non-supersymmetric, but it reduces for large $N$ to a supersymmetric 
one, namely ${\cal{N}}=4$ super Yang-Mills~\footnote{This theory has
  been discussed in Ref.~\cite{BFL9906} and its gravity dual has been
  constructed in Ref.~\cite{AA0003}.}.

One can extend the previous construction to a non-conformal theory by
considering the orbifold ${\mathbb{C}}^2 /{\mathbb{Z}}_2$ of the
previous orientifold. The gauge theory living in the world-volume of 
 $N$ fractional D3 branes of this orbifold is a non-supersymmetric one
 consisting of one gluon, two adjoint scalars and two Dirac fermions
 transforming according the two index (anti)symmetric representation
 of $SU(N)$. This gauge theory contains $4N^2$ bosonic and $4N(N\pm1)$
fermionic degrees of freedom. Its one-loop $\beta$-function is equal to:
\begin{eqnarray}
\beta (g_{YM} ) = \frac{g_{YM}^{3}}{(4 \pi)^2} \left[- \frac{11}{3}N +
2 \cdot \frac{N}{6} + 2 \cdot \frac{4}{3} \frac{N \pm 2}{2} \right] =
\frac{g_{YM}^{3}}{(4 \pi)^2} \left[- 2 N \pm \frac{8}{3} \right].
\label{beta2}
\end{eqnarray}
For large $N$ one recovers the Bose-Fermi degeneracy and the
$\beta$-function reduces to that of ${\cal{N}}=2$ super Yang-Mills.

Finally one can consider the orbifold $\frac{C^3}{(Z_2 \times Z_2) }$ 
of the previous orientifold. The gauge theory living in the world-volume of 
 $N$ fractional D3 branes of this orbifold is a non-supersymmetric one
 consisting of one gluon and one  Dirac fermion
 transforming according to the two-index (anti)symmetric representation
 of $SU(N)$. This gauge theory contains $2N^2$ bosonic and $2N(N\pm1)$
fermionic degrees of freedom. Its one-loop $\beta$-function is equal to:
\begin{eqnarray}
\beta (g_{YM} ) = \frac{g_{YM}^{3}}{(4 \pi)^2} \left[- \frac{11}{3}N 
+  \frac{4}{3} \frac{N \pm 2}{2} \right] =
\frac{g_{YM}^{3}}{(4 \pi)^2} \left[- 3 N \pm \frac{4}{3} \right] .
\label{beta3}
\end{eqnarray} 
For large $N$ one recovers the Bose-Fermi degeneracy and the
$\beta$-function reduces to that of ${\cal{N}}=1$ super Yang-Mills.
In conclusion we have constructed three systems of D3 branes that have
in their world-volume non-supersymmetric gauge theories that in the
large number of colours reduce to respectively ${\cal{N}}=1,2,4$ super
Yang-Mills.

\section{Gauge/gravity correspondence for the ``orientifold field theories''}
\label{sect4}

Let us consider the interaction between a stack of $N$
    D3 branes of the orientifold $ \frac{0B}{\Omega' I_6
    (-1)^{F_L}}$ and a D3 brane dressed with an external constant
    gauge field:
\begin{eqnarray}
Z^{o}=\int_{0}^{\infty} \frac{d \tau}{\tau} Tr_{{\rm
    NS-R}} \left[ \frac{e+\Omega'I_6}{2}\,
\frac{1+  (-1)^{F_s}}{2}\,(-1)^{G_{bc}}\,
%\frac{(-1)^{G_{\beta \gamma}} + (-1)^F }{2}  
P_{GSO} \,\,{ e}^{- 2 \pi \tau
L_0} \right] .
\label{annu92}
\end{eqnarray}
The term corresponding to $e$ gives the annulus diagram, while the other term
corresponds to the M{\"{o}}bius diagram. The annulus diagram does not give any
    contribution to the term quadratic in the gauge field as in
    ${\cal{N}}=4$ super Yang-Mills. The M{\"{o}}bius diagram gives instead 
a non-vanishing contribution that,
    in the open-string channel, is
    equal to ($ k = {e}^{- \pi \tau}$)
\begin{eqnarray}
\frac{1}{g^2_{\rm YM}} =\pm \frac{1}{(4\pi)^2} \int_{0}^{\infty}
\frac{d\tau}{\tau}
e^{-\frac{\rho^2\tau}{2\pi\alpha'} }
 \left(  \frac{f_2(ik)}{f_1(ik)}\right)^8 \left[\frac{1}{3\tau^2} +k
\frac{\partial}{\partial k} \log f_2^4(ik)\right] . 
\label{moe89}
\end{eqnarray}
The previous expression can be transformed in
    the closed-string channel by the transformation $\tau =
    \frac{1}{4t}$ ($q = {e}^{-\pi t}$):
\begin{eqnarray}
\frac{1}{g^2_{\rm YM}} =\pm \frac{1}{(4\pi)^2} \int_{0}^{\infty}
\frac{dt}{4t^3}
e^{-\frac{\rho^2}{8\pi\alpha't} }
\left( 
\frac{f_2(iq)}{f_1(iq)}\right)^8 \left[\frac{4}{3} + \frac{1}{
\pi}\partial_t \log f_2^4(iq)\right] .
\label{moe93}
\end{eqnarray}
They are finite both for
    $\tau \rightarrow \infty (t \rightarrow 0)$ and for $\tau \rightarrow
    0 (t \rightarrow \infty)$.
Hence there are no divergences at the string level if $\rho \neq 0$. 
Otherwise infrared divergences appear.
Performing the field-theory limit ($\alpha' \rightarrow 0$ and $
\sigma \equiv 2 \pi \alpha' \tau$ fixed) one gets UV divergences and
must introduce a cut-off ( $\mu = \frac{\rho}{2 \pi \alpha'}$):
\begin{eqnarray}
\frac{1}{g^2_{\rm YM} (\mu)}= 
\frac{1}{g^2_{\rm YM} (\Lambda )}
\pm \frac{1}{3 \pi^2}
\int_{\mu^2/\Lambda^2}^{\infty} \frac{d \sigma}{\sigma} 
e^{-\sigma } = \frac{1}{g^2_{\rm YM} (\Lambda )}
\mp \frac{1}{3\pi^2} \log
 \frac{\mu^2}{\Lambda^2} , 
\label{beta65}
\end{eqnarray}
that corresponds to the right $\beta$-function. The field theory limit in the
    closed-string channel ($t\rightarrow \infty$ and 
$\alpha'\rightarrow 0$ with $s=2\pi
\alpha't$ fixed) gives a vanishing contribution, $\underline{NOT}$
    reproducing the correct $\beta$-function of the theory.
This means that the gauge/gravity correspondence works only  in
    the large $N$ limit, where the contribution of the M{\"{o}}bius
    strip  is
    suppressed, the D3 branes are not interacting and  the gauge
    theory recovers the Bose-Fermi
    degeneracy in the spectrum.

Let us consider the orbifolds $C^2/ \mathbb{Z}_2$ 
of the previous orientifold. In this case we have two additional terms
with respect to the previous orientifold, namely we get:
\begin{eqnarray}
 \int_{0}^{\infty} \frac{d \tau}{\tau} Tr_{{\rm
    NS-R}} \left[ \left(\frac{1 +h}{2}\right) \left(\frac{e+\Omega'I_6}{2}
\right)\left( \frac{1+  (-1)^{F_s}}{2}\right)
  (-1)^{G_{bc}} P_{GSO}
%\left( \frac{(-1)^{G_{\beta \gamma}} + (-1)^F }{2} \right) 
\,\,{ e}^{- 2 \pi \tau L_0} \right].
\label{Z1orb}
\end{eqnarray}
Extracting the quadratic term in the gauge 
field we get (in the field
theory limit performed in the open string channel):
\begin{eqnarray}
\frac{1}{g_{YM}^{2}} = \frac{1}{16 \pi^2} \left[ 2N \mp
\frac{8}{3} \right] \log
  \frac{\mu^2}{\Lambda^2}
\label{runorior}
\end{eqnarray}
where the first term comes from the twisted annulus diagram and the second one
from the untwisted M{\"{o}}bius strip.
Extracting the running coupling constant from the
  closed string channel we get again only the leading term in $N$.
The twisted annulus and the twisted
    M{\"{o}}bius produce also a term proportional to the topological
    charge of the gauge theory, namely:
\begin{eqnarray}
-\frac{i(N \pm 2)}{32\pi^2}\int d^4x\, F_{\alpha\beta}^a {\tilde
F}^{a\,\alpha\beta} \int_{0}^\infty
\frac{d\tau}{\tau}
e^{-\frac{\rho^2 \tau}{2\pi\alpha'}} .
\nonumber
\end{eqnarray}
From it we can extract the value of
    $\theta_{YM}$ as before getting:
\begin{eqnarray}
\theta_{YM} =-i(N \pm 2)\int_{0}^\infty
\frac{d\tau}{\tau}
e^{-\frac{\rho^2 \tau}{2\pi\alpha'}} \rightarrow -2 \theta (N \pm 2).
\label{the93}
\end{eqnarray}
Since the $\theta_{YM}$-angle does not receive threshold corrections
    getting 
    contribution only from the massless open string states that, under
    open/closed string duality, are transformed only into massless
    closed string states, it is clear that it can be  equivalently obtained
    from either channels. The previous analysis can also be
    extended~\cite{DLMP0407}
    to the orbifold $\mathbb{C}^3/( \mathbb{Z}_2 \times \mathbb{Z}_2)$
    obtaining similar results.

%%%%%%%%%%%%%%%%%%%%%%%%%%%%%%%%%%%%%%%%%%% 
{\bf Acknowledgement} Work partially supported by the European
Community's Human Potential Programme under contract
MRTN-CT-2004-005104 `Constituents, fundamental forces and symmetries
of the universe' and by MIUR. 
%%%%%%%%%%%%%%%%%%%%%%%%%%%%%%%%%%%%%%%%%%

\end{document}